# Simplified Doppler frequency shift measurement enabled by Serrodyne optical frequency translation

Yang Chen[a,b,*] and Taixia Shi[a,b]

[a] *Shanghai Key Laboratory of Multidimensional Information Processing, East China Normal University, Shanghai 200241, China*
[b] *Engineering Center of SHMEC for Space Information and GNSS, East China Normal University, Shanghai 200241, China*
[*] ychen@ce.ecnu.edu.cn

**ABSTRACT**
A simplified Doppler frequency shift measurement approach based on Serrodyne optical frequency translation is reported. A sawtooth wave with an appropriate amplitude is sent to one phase modulation arm of a Mach-Zehnder modulator in conjunction with the transmitted signal to implement the Serrodyne optical frequency transition, as well as the optical phase modulation of the transmitted signal on the frequency-shifted optical carrier. The echo signal is applied to the other phase modulation arm of the Mach-Zehnder modulator. The optical signals from the two arms are combined in the Mach-Zehnder modulator, whose lower optical sidebands are selected by an optical bandpass filter and then detected in a photodetector. By simply measuring the frequency of the output low-frequency signal, the value and direction of DFS can be determined simultaneously. An experiment is performed. DFS from -100 to 100 kHz is measured for microwave signals from 6 to 17 GHz with a measurement error of less than ±0.03 Hz and a measurement stability of ±0.015 Hz in 30 minutes when a 500-kHz sawtooth wave is used as the reference.

## 1. Introduction

The relative motion between the signal transceiver and the reflector makes the received signal in the transceiver inconsistent with the transmitted signal in frequency. The frequency difference between the transmitted signal and the received signal is the Doppler frequency shift (DFS) [1]. Accurate identification and measurement of DFS plays an important role in many fields, for example, radar applications [2] and wireless communications [3]. Therefore, many methods have been reported to obtain the Doppler information of the echoes in the electrical domain [2, 4].

However, conventional electrical-based methods are limited in operating frequency, bandwidth, and tunability. To further meet the needs of future radar and communication systems to develop towards high-frequency bandwidth, DFS measurement is transferred to the optical domain and realized by microwave photonics [5]. There are three main DFS measurement methods based on microwave photonics: 1) employing a frequency shift module [6, 7]; 2) employing in-phase and quadrature (I/Q) detection [8, 9]; 3) employing a reference signal [10-15]. The methods based on adding a frequency shift module are commonly implemented by using an acousto-optic modulator and constructing parallel optical paths, which makes the system complicated. The methods based on I/Q detection need a pair of photodetectors (PDs) or a coherent receiver, and both the waveform and the signal frequency after I/Q detection need to be analyzed. The DFS measurement methods based on adding a reference signal reported earlier are complicated [10-13], which require at least two of the following three conditions: multiple modulators, high-frequency reference signal, and optical filtering. To simplify the system, two methods to realize DFS measurement is proposed in [14,15], which need either one or none of the conditions. However, even in the simplest method in [15], two Mach-Zehnder structures in an integrated modulator are used. Furthermore, 90º electrical hybrid couplers (90º HYB) are essential in the two methods [14,15], making the reference frequency difficult to be very low. Therefore, the frequency of the signal finally used to determine the DFS is with a relatively high frequency, resulting in an increased difficulty in identifying the signal.

In this paper, a simplified DFS measurement approach based on Serrodyne optical frequency translation is proposed. A sawtooth wave with an appropriate amplitude is sent to one phase modulation arm of a Mach-Zehnder modulator (MZM) in conjunction with the transmitted signal to implement the Serrodyne optical frequency transition, as well as the optical phase modulation of the transmitted signal on the frequency-shifted optical carrier. The echo signal is applied to the other phase modulation arm of the MZM. By selecting the lower optical sidebands of the optical signal from the MZM and detecting it in a PD, the value and direction of DFS can be determined by simply measuring the frequency of the output low-frequency signal from the PD. The sawtooth wave can be seen as a low-frequency reference, which shifts the optical carrier in one phase modulation arm of the MZM, so the proposed approach can be regarded as a method combining the frequency shift module and adding a reference signal mentioned above together.

## 2. Principle

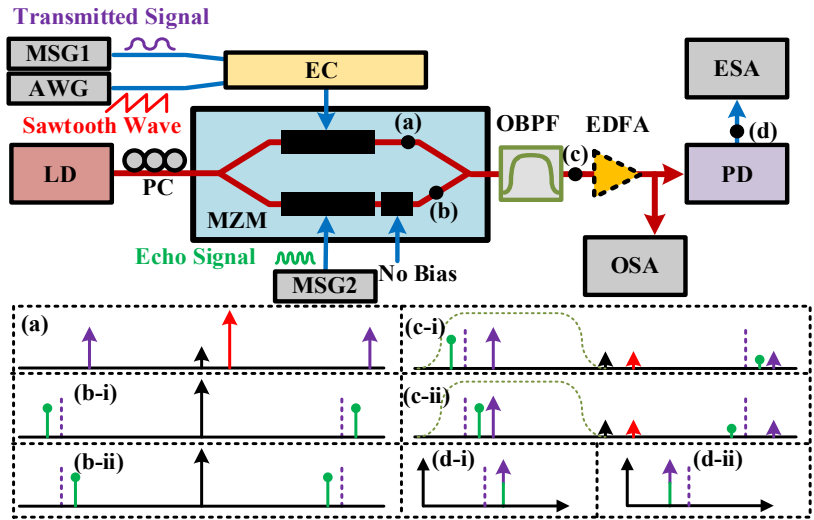

Fig. 1. Schematic diagram of the DFS measurement system. LD, laser diode; PC, polarization controller; MZM, Mach-Zehnder modulator; EC, electrical coupler; OBPF, optical bandpass filter; EDFA, erbium-doped fiber amplifier; PD, photodetector; MSG, microwave signal generator; AWG, arbitrary waveform generator; OSA, optical spectrum analyzer; ESA, electrical spectrum analyzer.

The schematic diagram of the DFS measurement system is shown in Fig. 1. An optical carrier generated from a laser diode (LD, ID Photonics CoBriteDX1-1-C-H01-FA) is sent to an MZM (Fujitsu FTM7937EZ200) via a polarization controller. The transmitted signal from a microwave signal generator (MSG1, Agilent 83630B) and a periodical sawtooth wave from an arbitrary waveform generator (AWG, Rigol DG2052) are combined in an electrical coupler (EC, TRM DMS285S, DC-18GHz, 6-dB insertion loss) and then sent to the upper phase modulation arm of the MZM. The amplitude of the sawtooth wave applied to the MZM is properly adjusted to make it equal to twice the half-wave voltage of the phase modulation arm [16] to implement the Serrodyne optical frequency translation. The frequency-shifted optical carrier by Serrodyne optical frequency translation is further phase-modulated by the transmitted signal, with the optical spectrum of the optical signal from the phase modulation arm shown in Fig. 1(a). The echo signal generated from another MSG (MSG2, HP 83752B) is directly sent to the other arm the MZM, and the optical spectra from this arm with positive and negative DFSs are shown in Fig. 1(b). The optical signal from the MZM is sent to an optical bandpass filter (OBPF, EXFO XTM-50) to select only the lower optical sidebands, as shown in Fig. 1(c). The filtered optical signal from the OBPF is amplified by an erbium-doped fiber amplifier (EDFA, Amonics AEDFA-PA-35-B-FA) and then detected in a PD (LSIPD-A75), with the spectra of the low-frequency signals from the PD shown in Fig. 1(d). It is noted that the EDFA is optional, which is mainly used when the echo signal power is very small. The reason why it is not used before the OBPF is to prevent the optical carrier from saturating the EDFA and occupying most of the output power of EDFA. By simply analyzing the frequency of the low-frequency signal from the PD, the value and direction of the DFS can be determined simultaneously.

Assuming the frequencies of the optical carrier, the transmitted signal, the echo signal, and the periodical sawtooth wave are $f_c$, $f_t$, $f_e$, and $f_r$, respectively, the frequencies of the lower sidebands in Fig. 1(a) and (b) are $f_c+f_r-f_t$ and $f_c-f_e$. Thus, the frequency of the generated low-frequency signal shown in Fig. 1(d) is $f=f_r-f_t+f_e$. Commonly, $f_r$ is selected to be a very low frequency, i.e., $f_r \ll f_t, f_e$, so the generated signal from the PD is easier to be analyzed. By simply obtaining the frequency of the low-frequency signal from the PD, the DFS $f_d$ can be calculated as $f_d = f_e - f_t = f - f_r$, in which both the value and direction of the DFS are included.

## 3. Experimental results and discussion

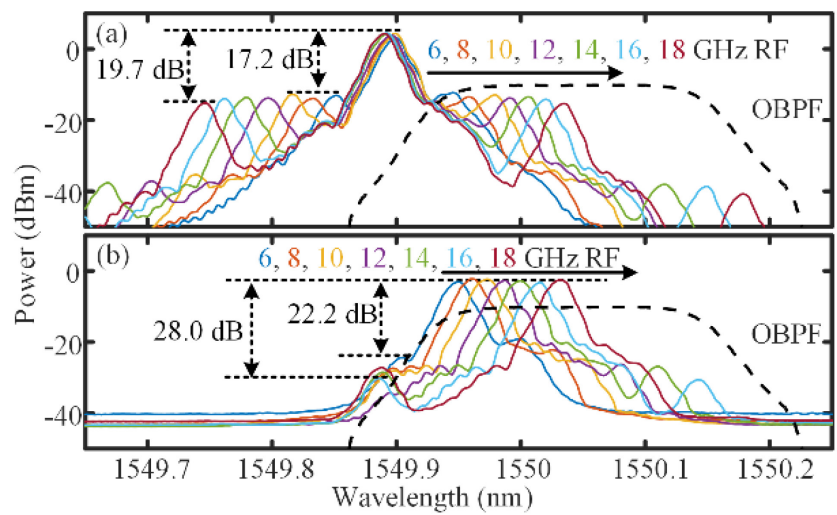

Fig. 2. Optical spectra at the outputs of (a) the MZM and (b) the EDFA. The dotted lines represent the filtering response of the OBPF.

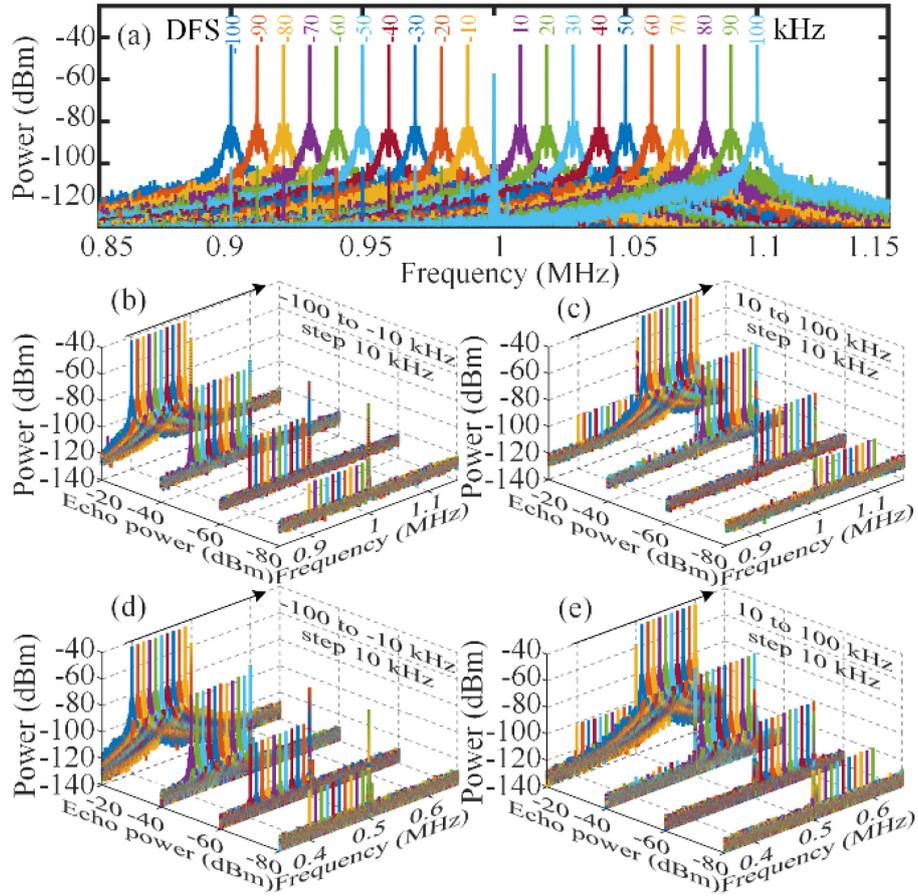

Fig. 3. Electrical spectra of the low-frequency signal from the PD (a) with 6-GHz transmitted signal, 1-MHz sawtooth wave, and -20 dBm echo power, (b), (c) with 10-GHz transmitted signal, 1-MHz sawtooth wave, and -20 to -80 dBm echo power, (d), (e) with 17-GHz transmitted signal, 500-kHz sawtooth wave, and -20 to -80 dBm echo power.

An experiment is performed based on the setup shown in Fig. 1. The power and wavelength of the optical carrier are set to 16 dBm and 1549.895 nm, respectively. A sawtooth wave from the AWG with a frequency of 1MHz and an amplitude of 14.95 V is combined with a 5-dBm transmitted signal from MSG1. The power of the echo signal from MSG2 is set to 0 dBm. The amplitude of the sawtooth wave is carefully selected to implement the best Serrodyne optical frequency translation. The center wavelength and bandwidth of the OBPF are set to 1550.040 nm and 24.85 GHz, respectively. Fig. 2 shows the optical spectra at the outputs of the MZM and the EDFA captured by an optical spectrum analyzer (Ando, AQ6317B). The frequencies of the transmitted and echo signals are tuned from 6 to 18 GHz with a DFS of -100 kHz. As can be seen, the carrier is always filtered out with a suppression ratio of more than 22.2 dB when the frequency of the transmitted and echo signal changes. After the OBPF, whose filtering response is shown in dotted lines, only the lower optical sidebands from the MZM are reserved. However, the Serrodyne optical frequency translation cannot be observed from the optical spectrum due to the limited resolution of OSA.

Then, the electrical spectra from the PD are shown in Fig. 3 by using an electrical spectrum analyzer (ESA, R&S FSV4). When a 6-GHz transmitted signal, a 1-MHz sawtooth wave, and a -20-dBm echo signal are applied to the system, the electrical spectrum is shown in Fig. 3(a). Besides the desired low-frequency signal, a suppressed single-tone signal is fixed at 1 MHz. Furthermore, some spurs with very low power are also generated. The 1-MHz undesired signal is mainly generated by beating the residual original optical carrier and the shifted optical carrier, whereas the low-power spurs are mainly generated due to the unideal sawtooth wave. By tuning the sawtooth wave amplitude, the power of the spurs varies. The sawtooth wave amplitude is carefully adjusted to implement the best Serrodyne optical frequency translation with the lowest spurs. Fig. 3(b) and (c) show the electrical spectra when the transmitted signal is tuned to 10 GHz and the echo power changes from -20 to -80 dBm. As can be seen, the 1-MHz signal remains unchanged because it has nothing to do with echo power. The low-power spurs are reduced together with the desired low-frequency signal as it is generated from the beating between the echo sideband and the undesired sideband caused by non-ideal Serrodyne optical frequency translation. Fig. 3(d) and (e) show the electrical spectrum when the transmitted signal is tuned to 17 GHz and the echo power changes from -20 to -80 dBm. In this case, the frequency of the sawtooth wave is changed to 500 kHz. Similar results are obtained except that the signal center is shifted from 1 MHz to 500 kHz.

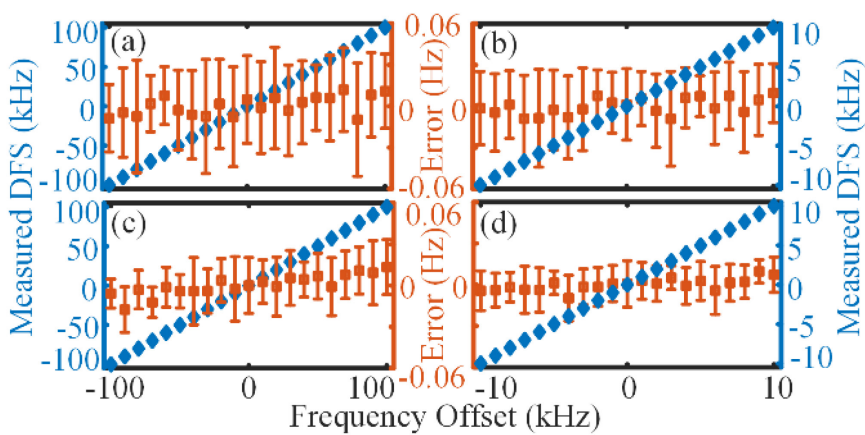

Fig. 4. Measured DFSs and measurement errors when the frequency of the transmitted signal is 17 GHz, (a) the DFS is from -100 to 100 kHz, the frequency of the sawtooth wave is 1 MHz, (b) the DFS is from -10 to 10 kHz, the frequency of the sawtooth wave is 1 MHz, (c) the DFS is from -100 to 100 kHz, the frequency of the sawtooth wave is 500 kHz, (d) the DFS is from -10 to 10 kHz, the frequency of the sawtooth wave is 500 kHz.

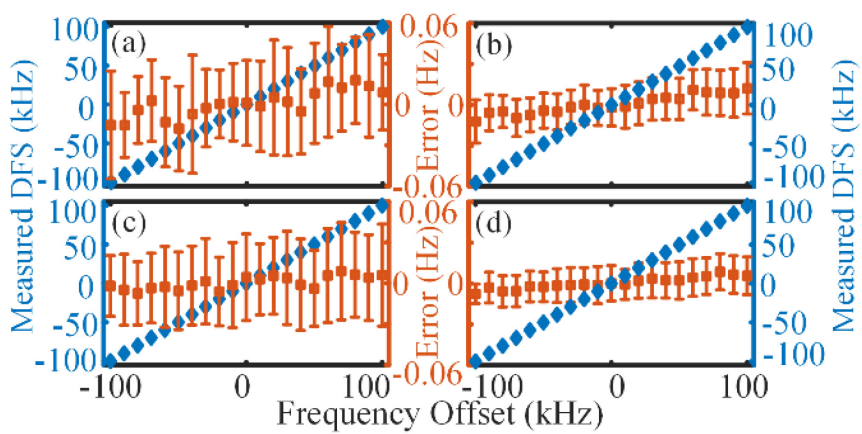

Fig. 5. Measured DFSs and measurement errors when DFS is from -100 to 100 kHz, (a) the frequency of the transmitted signal is 10 GHz, the frequency of the sawtooth wave is 1 MHz, (b) the frequency of the transmitted signal is 10 GHz, the frequency of the sawtooth wave is 500 kHz, (c) the frequency of the transmitted signal is 6 GHz, the frequency of the sawtooth wave is 1 MHz, (d) the frequency of the transmitted signal is 6 GHz, the frequency of the sawtooth wave is 500 kHz.

Fig. 4 shows the DFS measurement errors when the frequency of the transmitted signal is 17 GHz and the frequency of the periodical sawtooth wave is 500 kHz or 1 MHz. In the experiment, each DFS is measured five times with an interval of 10 minutes. The mean DFS values of the five measurements are represented by blue diamonds, and the standard deviations of the DFS errors are represented by the red error bars. Fig. 4(a) and (b) show the case with a 1-MHz sawtooth wave. The DFS is accurately measured with a measurement error of less than ±0.05 Hz. When the frequency of the sawtooth wave is changed to 500 kHz, the standard deviations of the DFS errors decrease as shown in Fig. 4(c) and (d). In this case, the measurement error is approximately less than ±0.03 Hz. Fig. 5 shows the DFS measurement errors when the frequency of the transmitted signal is adjusted to 6 and 10 GHz. It can be seen that the DFS measurement performance does not have significant variations when the frequency of the transmitted signal changes. From Figs. 4 and 5, it can be further observed that the lower the frequency of the desired signal from the PD, the lower the measurement error in the same system. This is also a reason why it is desirable to measure the DFS in a lower frequency band in addition to the fact that low-frequency signals are easier to measure.

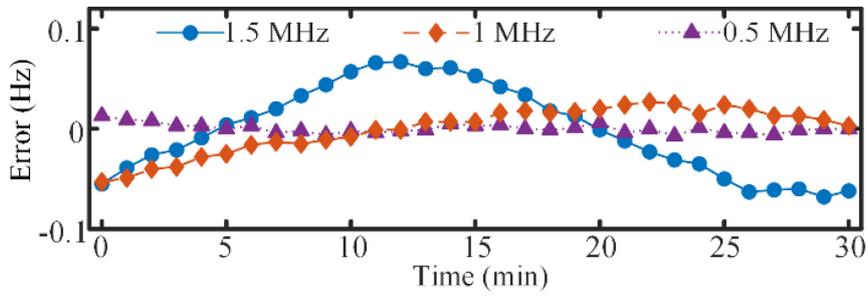

Fig. 6. The stability of the measurement in 30 minutes when the DFS is 100 kHz and the frequency of the sawtooth wave is 500 kHz, 1 MHz, or 1.5 MHz, respectively.

Fig. 6 shows the stability of the measurement in 30 minutes. In the experiment, the frequency of the transmitted signal and the DFS is fixed to 17 GHz and 100 kHz. The frequency of the sawtooth wave is respectively set to 500 kHz, 1 MHz, and 1.5 MHz, and the stability is measured one by one. In each case, the DFS is measured every 1 minute. As can be seen, within 30 minutes, the measurement errors of the three cases have a small difference. When the frequency of the sawtooth wave is lower, better measurement stability can be achieved. For example, the stability is within ±0.015 Hz when the frequency of the sawtooth wave is 500 kHz. In comparison, the stability decreases to around ±0.05 Hz when the frequency of the sawtooth wave is 1.5 MHz. In fact, in the experiment, MSG2 is synchronized with MSG1 by using the 10-MHz reference from MSG1, while the AWG operates using its internal clock. As discussed in [15], the reference is the key to achieve good long-term stability because the long-term frequency drifts of the transmitted signal and the echo cancel each other. However, when the measurement error is very small, the residual frequency drift difference between the transmitted signal and the echo is close to the stability of reference. In this case, the drift difference also affects the stability.

The tunability of the system is mainly limited by the OBPF, which is used to select optical sidebands on one side. Therefore, when XTM-50 OBPF is used, the lower bound of measurement range is limited to around 6 GHz. In the experiment, the system is demonstrated from 6 to 17 GHz, and the upper bound of the measurement range is mainly limited by the EC and MSG2. To further increase the upper bound of the measurement range, the bandwidth of the OBPF needs to be further increased and EC with a wider operating bandwidth needs to be used desired.

Table 1. Comparison of DFS measurement methods using a low-frequency reference

|  | Demonstrated Bandwidth(GHz)/Error(Hz) | Modulator Structures/ Optical Filter | Hybrid Coupler/ Phase Shifter | Reference Frequency Used |
|---|---|---|---|---|
| [13] | 12-18/±0.1 | 3/1 | 0/1 | 300 kHz |
| [14] | 7-16/±0.05 | 3/0 | 3/0 | 1.8 GHz |
| [15] | 6.9-16.1/±0.04 | 2/0 | 1/0 | 1.7 GHz |
| This work | 6-17/±0.03 [a] | 1/1 | 0/0 | 500 kHz [b] |

[a] The error is obtained by using a 500-kHz sawtooth wave; [b] the reference frequency can be much lower than 500 kHz according to the DFS.

In a DFS measurement system, the signal used to determine the DFS is better at very low frequencies. However, in the reported method using a low-frequency reference [14,15], the 90º HYB severely limits the lower bound of the reference frequency that can be used because the reference and the transmitted signal are sent to the 90º HYB simultaneously to simplify the system by using fewer modulator structures. Although the method in [13] employs a reference at 300 kHz, the reference, the transmitted signal, and the echo are sent to the system independently, resulting in a complicated structure with three modulator structures. Thanks to the Serrodyne optical frequency translation, the 90º HYB is replaced by an EC, which can easily work from DC to a very high-frequency band. Thus, without increasing the number of modulator structures, the signal from the PD is greatly reduced to any low frequency from DC, which is much easier to be measured. Furthermore, the system employs only a single MZM with two modulation arms. As is well known, the Serrodyne optical frequency translation is ideal only when the sawtooth wave is ideal. Otherwise, the Serrodyne optical frequency translation is realized with additional multiple spectral components with a frequency spacing equal to $f_r$. With the increase of the ratio of the fall time to the period of the sawtooth wave, the power levels of the additional sidebands increase [16]. Fortunately, for DFS measurement, the frequency of the sawtooth wave is desired to be only greater than twice the maximum DFS, for example, 500 kHz or 1 MHz in this paper, which is easier to be generated with a smaller ratio of the fall time to the period, resulting in very small undesired sidebands and very small influence on the DFS measurement. The frequency of the sawtooth wave can be further decreased to be much lower than 500 kHz according to the DFS to be measured. A detailed comparison between this work and other works using a low-frequency reference is given in Table 1.

## 4. Conclusions
In summary, a DFS measurement method based on Serrodyne optical frequency translation is proposed and verified. The key significance of the work is that the frequency of the low-frequency reference is reduced to 500 kHz, which is close to the general value of DFS, under the premise of using a single Mach-Zehnder structure. The system features an ultra-simple structure, wide operating bandwidth, good tunability, and ultra-low-frequency reference. Experimental results show that DFS from -100 to 100 kHz is measured for microwave signals from 6 to 17 GHz with a measurement error of less than ±0.03 Hz and a measurement stability of ±0.015 Hz in 30 minutes when a 500-kHz sawtooth wave is used as the reference. The proposed method is a promising solution to implement DFS measurement in the optical domain, which can provide a feasible solution for DFS measurement in radar, communication, and other systems.

**Funding**

National Natural Science Foundation of China (NSFC) (61971193); Natural Science Foundation of Shanghai (20ZR1416100); Open Fund of State Key Laboratory of Advanced Optical Communication Systems and Networks, Peking University, China (2020GZKF005); Science and Technology Commission of Shanghai Municipality (18DZ2270800).

**Conflicts of interest**

The authors declare no conflicts of interest.

**References**

1. T. P. Gill, The Doppler effect: An introduction to the theory of the effect. (Logos/Academic, 1965.)
2. V. C. Chen, F. Li, S. Ho, and H. Wechsler, "Micro-Doppler effect in radar: Phenomenon, model, and simulation study," IEEE Trans. Aerosp. Electron. Syst. **42**(1), 2-21 (2006).
3. A. Salberg and A. Swami, "Doppler and frequency-offset synchronization in wideband OFDM," IEEE T. Wirel. Commun. **4**(6), 2870-2881 (2005).
4. L. Yang, G. Ren, and Z. Qiu, "A novel doppler frequency offset estimation method for DVB-T system in HST environment," IEEE Trans. Broadcast. **58**(1), 139-143 (2012).
5. X. Zou, B. Lu, W. Pan, L. Yan, A. Stöhr, and J. Yao, "Photonics for microwave measurements," Laser Photon. Rev. **10**(5), 711-734 (2016).
6. X. Zou, W. Li, B. Lu, W. Pan, L. Yan, and L. Shao, "Photonic approach to wide-frequency-range high-resolution microwave/millimeter-wave Doppler frequency shift estimation," IEEE Trans. Microw. Theory Tech. **63**(4), 1421-1430 (2015).
7. C. Yi, H. Chi, B. Yang, and T. Jin, "A PM-based approach for Doppler frequency shift measurement and direction discrimination," Opt. Commun. **458**, 124796 (2020).
8. F. Zhang, J. Shi, and S. Pan, "Photonics-based wideband Doppler frequency shift measurement by in-phase and quadrature detection," Electron. Lett. **54**(11), 708-710 (2018).
9. P. Li, L. Yan, J. Ye, X. Feng, W. Pan, B. Luo, X. Zou, T. Zhou, and Z. Chen, "Photonic approach for simultaneous measurements of Doppler-frequency-shift and angle-of-arrival of microwave signals," Opt. Exp. **27**(6), 8709-8716 (2019).
10. L. Xu, Y. Yu, H. Tang, and X. Zhang, "A simplified photonic approach to measuring the microwave Doppler frequency shift," IEEE Photon. Technol. Lett. **30**(3), 246-249 (2018).
11. C. Huang, H. Chen, and E. H. W. Chan, "Simple photonics-based system for Doppler frequency shift and angle of arrival measurement," Opt. Exp. **28**(9)**,** 14028-14037 (2020).
12. Z. Cui, Z. Tang, S. Li, Z. He, and S. Pan, "On-chip photonic method for Doppler frequency shift measurement," in 2019 International Topical Meeting on Microwave Photonics (MWP, 2019), p. 1-3.
13. C. Huang, E. H. W. Chan, and C. B. Albert, "Wideband DFS measurement using a low-frequency reference signal," IEEE Photon. Technol. Lett. **31**(20), 1643-1646 (2019).

14. P. Zuo and Y. Chen, "Photonic-assisted filter-free microwave Doppler frequency shift measurement using a fixed low-frequency reference signal," J. Light. Technol. **38**(16), 4333-4340 (2020).
15. Y. Chen, P. Zuo, T. Shi, and Y. Chen, "Photonic-enabled Doppler frequency shift measurement for weak echo signals based on optical single-sideband mixing using a fixed low-frequency reference," J. Light. Technol. **39**(10), 3121-3129 (2021).
16. L. M. Johnson and C. H. Cox, "Serrodyne optical frequency translation with high sideband suppression," J. Light. Technol. **6**(1), 109-112 (1988).